\documentclass[%
preprint, superscriptaddress,
 amsmath,amssymb, aps, onecolumn,
]{revtex4-1} 
\usepackage{graphicx} 
\usepackage{subcaption}
\usepackage{color}
\begin{document} 

\title{Simulations to study the static polarization limit for RHIC
lattice with the Polymorphic Tracking Code}

\author{Zhe~Duan} 
\affiliation{Key
Laboratory of Particle Acceleration Physics and Technology, Institute of High
Energy Physics, Chinese Academy of Sciences, 100049 Beijing, China}
\affiliation{University of Chinese Academy of Sciences, Beijing 100049, China}

\author{Qing~Qin} 
\affiliation{Key Laboratory of Particle
Acceleration Physics and Technology, Institute of High Energy Physics, Chinese
Academy of Sciences, 100049 Beijing, China}

\date{\today}

\begin{abstract} 
    We report a study of spin dynamics based on simulations with the Polymorphic Tracking
    Code~(PTC),
    exploring the dependence of the static polarization limit on various
    beam parameters and lattice settings for a practical RHIC lattice.
\end{abstract}
\maketitle

\section*{Introduction}

    The motion of the spin expectation value (the ``spin'') $\vec{S}$ of a 
charged particle traveling in the electric and magnetic fields in a circular accelerator is described by the Thomas-BMT equation~\cite{L.H.Thomas1927}, %
${d\vec{S}}/{d\theta}=\vec{\Omega}(\vec{z}, \theta)\times\vec{S}$,
where $\theta$ is the azimuthal angle, $\vec{z}$ is the location in the 
6-dimensional (6D) phase space  and 
$\vec{\Omega}$ contains the electric and magnetic fields in the laboratory frame.
For calculations it is convenient to write 
$\vec{\Omega}(\vec{z}, \theta) = {\vec{\Omega}}_0(\theta) + \vec{\omega}(\vec{z}, \theta)$
where ${\vec{\Omega}}_0(\theta)$ is the contribution from motion on the closed orbit and $\vec{\omega}(\vec{z}, \theta)$ is
the contribution from the synchro-betatron motion.

It is often necessary to describe spin motion with the help 
of a unit vector field
${\hat n}(z,\theta)$ (the ''invariant spin field'', or ISF for short)
\cite{D.P.Barber2004} and this object will play a central r{\^o}le in this paper.
This satisfies the Thomas-BMT equation along particle trajectories and is periodic with respect to
$\theta$:
$\vec{n}(\vec{z},\theta+2\pi)=\vec{n}(\vec{z},\theta)$. 
The product $J_s=\vec{S}\cdot\vec{n}$ is an invariant of motion, and the motion of
$\vec{S}$ is simply a precession around the local $\vec{n}$-axis.
The spin precession frequency
around $\hat{n}$ is characterized by the amplitude dependent spin tune
$\nu_s$~\cite{D.P.Barber2004}.
Let us assume that the 6D orbital motion is integrable and away from orbital resonances and spin-orbit resonances (defined below). Then  the static
polarization limit~\cite{Vogt} $P_{\text{lim}}=\vert\langle\vec{n}(\vec{z},\theta)\rangle\vert$, with the inner average
taken over orbital phases,
is the maximum achievable equilibrium beam polarization on a phase-space torus.
On the closed orbit, $\vec{n}$ is denoted by $\vec{n}_0$ and it is normally vertical
in the arcs and  $\nu_s$ reduces to the closed orbit spin tune $\nu_0$. 
In a perfectly aligned planar ring, $\nu_0=G\gamma_0$, 
where $G=1.79284739$ for protons, and $\gamma_0$ is
the relativistic factor for the design energy.  
The $\vec{n}$-axis diverges from $\vec{n}_0$ near
the following spin-orbit resonances (or ``spin resonances'' in short), and $P_{\text{lim}}$ 
can be small,
\begin{equation}
    \label{spin resonances} \nu_s=k+k_x\nu_x+k_y\nu_y+k_z\nu_z, ~ ~ ~
    k,~k_x,~k_y,~k_z\in\mathbb{Z}.
\end{equation}
Note that $\nu_s$ is usually undefined on orbital resonances,
otherwise $\nu_s$ is only a function of the
orbital actions $\vec{J}$ and the optical state of the ring. In particular, if $\nu_s(\vec{J})$ is an amplitude dependent spin tune,
then the fractional part of $\pm\nu_s(\vec{J})+l_0+\vec{l}\cdot\vec{\nu}$, with $l_0\in\mathbb{Z}$ and $\vec{l}\in\mathbb{Z}^3$ is also an
amplitude dependent spin tune. In other words, there is an equivalence class of
amplitude dependent spin tunes~\cite{D.P.Barber2004}. In addition, the dip in $P_{\text{lim}}$ across a spin resonance
is accompanied by a jump in the amplitude dependent spin tune so that a system can never 
actually sit at a resonance as defined in Eq. (1)~\cite{Vogt, Hoffstaetter:2004ra}. 
The order of a resonance is defined as $|k_x| + |k_y| + |k_z|$. 
Normally $\nu_s$ stays close  to $\nu_0$. 
 
In a planar ring, the most important spin resonances are those due to 
vertical closed orbit distortions, which occur near $\nu_0=k,~ k\in\mathbb{Z}$,
namely the imperfection spin resonances; and those driven by the vertical
betatron oscillations, which occur near $\nu_0=k\pm\nu_y,~ k\in\mathbb{Z}$,
namely the first order intrinsic spin resonances with $|k_y|= 1$.
The major challenge in a high energy polarized proton
synchrotron like RHIC is to preserve the beam polarization during acceleration~\cite{RHICPPCMan, prl_205GeV}.
The well-known Frossart-Stora formula~\cite{Froissart-Stora} describes the
polarization loss after crossing a single isolated spin resonance. 
Introduction of a pair of diametrically opposed orthogonal Siberian snakes~\cite{Derbenev1978} renders the closed 
orbit spin tune to be $\nu_0=0.5$ and independent of the beam
energy. Therefore intrinsic resonances are avoided for normal
$\nu_y$ during acceleration and even with misalignments,
$\nu_0$ remains close to $0.5$ independently of the energy so
that imperfection resonances are avoided too.
However, 
at rational vertical tunes 
satisfying the condition $1/2+k=m\nu_y,~  m,k\in\mathbb{Z}$,
there still can be strong loss of polarization during acceleration. This phenomeon is traditionally called 
 ``snake resonance''~\cite{S.Y.Lee1986, S.Y.Lee1997} although for rational $\nu_y$ the amplitude dependent 
 spin tune does not exist so that this condition does not correspond to higher-order resonance as defined in Eq.~\ref{spin resonances} . 
Of course in a real ring with misalignments, $\nu_0$ need not be exactly 0.5. Then pairs of real resonances 
in the sense of Eq.~\ref{spin resonances} with irrational $\nu_y$ can appear, sitting symmetrically on each side of the  $\nu_y$
for the ``snake resonance''. These doublets can also cause loss of polarization during acceleration. 

Inspired by the Froissart-Stora formula, in rings without snakes, it is common practice to compute the strengths of the imperfection
and intrinsic resonances for a practical lattice, and identify the most dangerous ones. Near one of these 
dangerous spin resonances, one can invoke  the so-called ``single resonance model''~\cite{Courant1980, S.R.Mane1988}.
This model is based on the rotating wave approximation whereby the effect of 
$\vec{\omega}$ is dominated by one particular Fourier harmonic
at $\kappa=k+k_x\nu_x+k_y\nu_y+k_z\nu_z,~k,k_x,k_y,k_z\in\mathbb{Z}$, 
with $|k_x|+ |k_y|+|k_z|\leq1$, and the corresponding resonance strength is $\epsilon_{\kappa}$.
This lattice-independent model is analytically solvable, and can be extended to include Siberian snakes, often modeled 
by ``point-like'' spin rotations. 
These lattice-independent models have been extensively studied for proton storage rings.
Analytical solutions for ${\nu_s}$ \cite{S.R.Mane2008} and the $\vec{n}$-axis
\cite{S.R.Mane2002b, S.R.Mane2003, S.R.Mane2008} have been obtained, as well as
$P_{\text{lim}}$ \cite{S.R.Mane2004}. Thus as in Ref.~\cite{S.R.Mane2008}, 
$\nu_s$ is 0.5 independently of the betatron amplitude, 
if $\nu_y$ is irrational.
Moreover,  as shown in
\cite{Barber2003, Barber2006} simulations
with these lattice-independent models are a great help for understanding the
peculiar features of spin dynamics for rational $\nu_y$ and, in particular, on and near ``snake resonances''. 

However, in high-energy proton rings, the basic spin resonances (i.e., without snakes) of interest might not be well isolated,
then a lattice-dependent study is necessary. For example, as we explain in Sections~\ref{sec:amp} 
and ~\ref{sec:tune}  below, 
doublets of higher-order resonances at irrational $\nu_y$  can occur 
near to the $\nu_y$ of a ``snake resonance''
since $\nu_s$ need not be 0.5. 

An extensive lattice-dependent study of the behavior 
of $P_{\text{lim}}$ throughout the whole energy range was made in the
study of polarized proton beams up to 920~GeV in HERA~\cite{Vogt,
G.H.Hoffstaetter2006}, as an approach complementary to direct spin tracking for acceleration.
It is also interesting to study $P_{\text{lim}}$ for the store conditions
of RHIC, with a beam energy of 255GeV, since this is relevant for the study of polarization 
variation during physics stores with constant beam energy~\cite{V.Schoefer2012, Spin2014}.

In section~\ref{sec:model}, methods for modeling and simulation based on the Polymorphic Tracking Code~(PTC)
are described, and in section~\ref{sec:simulation} the results of the simulation are presented.

\section{The model for the simulations and the methods\label{sec:model}}

The simulations in this paper utilize the Polymorphic Tracking Code~(PTC)
developed by E. Forest \cite{Schmidt:2002vp}. Designed to model various
geometries of particle accelerators, PTC is capable of symplectic tracking of
the orbital motion and length-preserving transport of spin \cite{S.Mane2009PTC},
where vectors of particle
coordinates and Taylor maps can be tracked in a polymorphic manner, and where the
latter enables the normal form analysis of the one-turn map using FPP
\cite{fppipac2006}.  Fortran programs have been developed to do the spin
tracking, which call PTC as a library. The modeling of the RHIC lattice is
presented first, followed by an explanation of the methods of simulation.  

\subsection{\label{3.1} Modeling of the RHIC lattice}

The MADX model of the RHIC lattice is exported into an input file for PTC, which is 
read by the Fortran program.
When a particle is tracked through an integration step of a magnet body, 
the orbital transfer map is sandwiched in between two spin kicks in equal amounts and 
each orbital transfer map is a second-order symplectic integrator 
while  each spin kick is represented by  a $3\times3$ orthogonal rotation matrix. 
The quadrupoles must be split into many integration steps to ensure the accuracy
of orbital and spin tracking. An upper limit is set for the spin rotation angle of each integration
step, calculated for the betatron amplitude of the tracked particle. In this study, we use about 7 integration
steps for each arc quadrupole, and up to 81 integration steps for the quadrupoles in the final focus
triplets.

There are two different implementations of a Siberian snake in this 
study. The first method implements a zero length spin kick that rotates a
spin by 180 degrees around an axis in the lattice, namely a ``point-like
snake'', while the second method implements helical dipoles~\cite{HelicalDipole} into the lattice,
namely a ``helical dipole snake''.
In PTC, a helical dipole is modeled with a symplectic transfer map accurate to the 4-th order. Note 
that the longitudinal magnetic field inside helical dipoles will introduce a small transverse coupling
at large orbital excursions.

This study implements a  pair of diametrically opposed 
 snakes
and therefore the closed-orbit spin tune is 0.5.
Note that the spin rotators around IP6 (interaction point at 6 o'clock)
and IP8 (interaction point at 8 o'clock) 
are not included in 
this study, but they can also be modeled in a similar way.

The beam-beam interaction is the major beam-current-dependent effect that might affect
the beam polarization during physics stores. 
The effect of the beam-beam interaction on the beam polarization 
was studied through long-term tracking, with a lattice-independent model~\cite{Batygin} and 
element-by-element tracking~\cite{Lucio1997}, where the beam-beam kick on spin motion was
taken into account.
However, the total beam-beam parameter~\cite{Hirata} from the two IPs in RHIC is within
-0.015, and the contribution from the linearized beam-beam kick to the intrinsic resonance strength,
is less than that of an arc quadrupole~\cite{Lucio1997}. Therefore, the beam-beam spin kick is not too important. 
Nevertheless, the beam-beam interaction also
introduces an incoherent tune shift for beam particles. This effect is studied in this work,
with a thin-lens weak-strong beam-beam kick implemented for the orbital motion.

\subsection{Simulation method}

The $\vec{n}$-axis is calculated in PTC using
stroboscopic averaging~\cite{K.Heinemann1996}. 
Once $\vec{n}$ is computed for 
a phase space point $\vec{z}$ at an azimuth $\theta$, a particle is launched at
the same location with spin parallel to $\vec{n}$ and tracked for $5000$ turns. 
If none of the three orbital tunes is rational, then the turn-by-turn orbital
coordinates trace phase space points on the same torus, and the turn-by-turn spins $\{\vec{S}^{\parallel}_{j}\}, j=1,2{\cdots}5000$
are the local $\vec{n}$ at the corresponding phase space points.
Therefore, $P_{\text{lim}}$ can be calculated as an average over $\{\vec{S}^{\parallel}_{j}\}$ 
for such a phase space torus.  
In addition, the amplitude dependent spin tune can be
obtained by a Fourier analysis of the spin motion according to Ref.~\cite{D.P.Barber2004}.
So a particle is launched with its spin perpendicular to $\vec{n}$, and tracked for $5000$ turns,
and a turn-by-turn spin series $\{\vec{S}^{\bot}_{j}\}$ is obtained. 
The NAFF algorithm \cite{Laskar}
is then applied to calculate the fundamental frequency of the complex series 
$\{S^{\bot}_{xj}+iS^{\bot}_{zj}\}$, i.e., the amplitude dependent spin tune, 
in the range $[0.4,0.5]$. 
 
\section{Simulation of RHIC polarization at store\label{sec:simulation}}
In this section, the dependence of $P_\text{lim}$
on various beam parameters is studied for the Run 12 baseline lattice of the RHIC Blue ring,
during physics stores. The default working point is $\nu_x/\nu_y=28.695/29.685$, and 
$\beta_x/\beta_y=0.62\text{m}/0.64\text{m}$ at the two interaction points IP6 and IP8
where the detectors STAR and PHENIX are located, respectively.
the chromaticities are corrected to $\xi_x/\xi_y=1.85/2.27$ with two families of arc sextupoles.
$\vec{n}$ and $P_{\text{lim}}$ are calculated at IP6, while 
$\nu_s$ is independent of the azimuthal angle.
The betatron amplitudes are normalized with the amplitude corresponding
to a normalized 95\% emittance of $10\pi\text{mm}\cdot\text{mrad}$. Two cases with two different
beam energies are simulated, and the relevant parameters are shown in
Table.~\ref{table:RHIC}. Note that Case 1 is the beam energy of physics data taking,
while Case 2 corresponds to the beam energy of a very strong intrinsic resonance around 
$[\nu_y]=0.7$ during the energy ramp. 
\begin{table}[htb]
   \centering
   \begin{ruledtabular}
   \begin{tabular}{lcc}
     Parameter             & Case 1    &  Case 2    \\ 
     \colrule
     Beam energy(GeV)      & 254.8675   &  200.6022  \\
     $G\gamma_0$             & 487        &  383.31     \\
     Normalized 95\% emittance($\pi~\text{mm}\cdot\text{mrad}$)  & 10.0  & 10.0 \\
     Intrinsic resonance strength&   0.0026/0.0013\footnote{for $G\gamma_0=486.69/487.31$, respectively.}       &   0.175             \\ 
   \end{tabular}
   \end{ruledtabular}
   \caption{Parameters of the RHIC lattice used in scans of $P_\text{lim}$} 
   \label{table:RHIC}
 \end{table}

\subsection{Amplitude scan \label{sec:amp}}
For a fixed lattice, different betatron amplitudes contribute to different
underlying spin resonance strengths. Therefore they will lead to different
behaviors of the $\vec{n}$-axis on the tori and different values of
$P_{\text{lim}}$. In this study, a scan of $P_{\text{lim}}$ and $\nu_s$ over
different vertical betatron amplitudes is investigated for Case 1 and Case 2,
while the synchrotron amplitude is set to zero. Note that for the cases with 
helical dipole snakes,
the snakes introduce a small transverse coupling so that there is a nonzero but small horizontal betatron amplitude.

\begin{figure}[htb]
    \begin{subfigure}{0.5\linewidth}
        \centering
        \includegraphics[width=75mm]{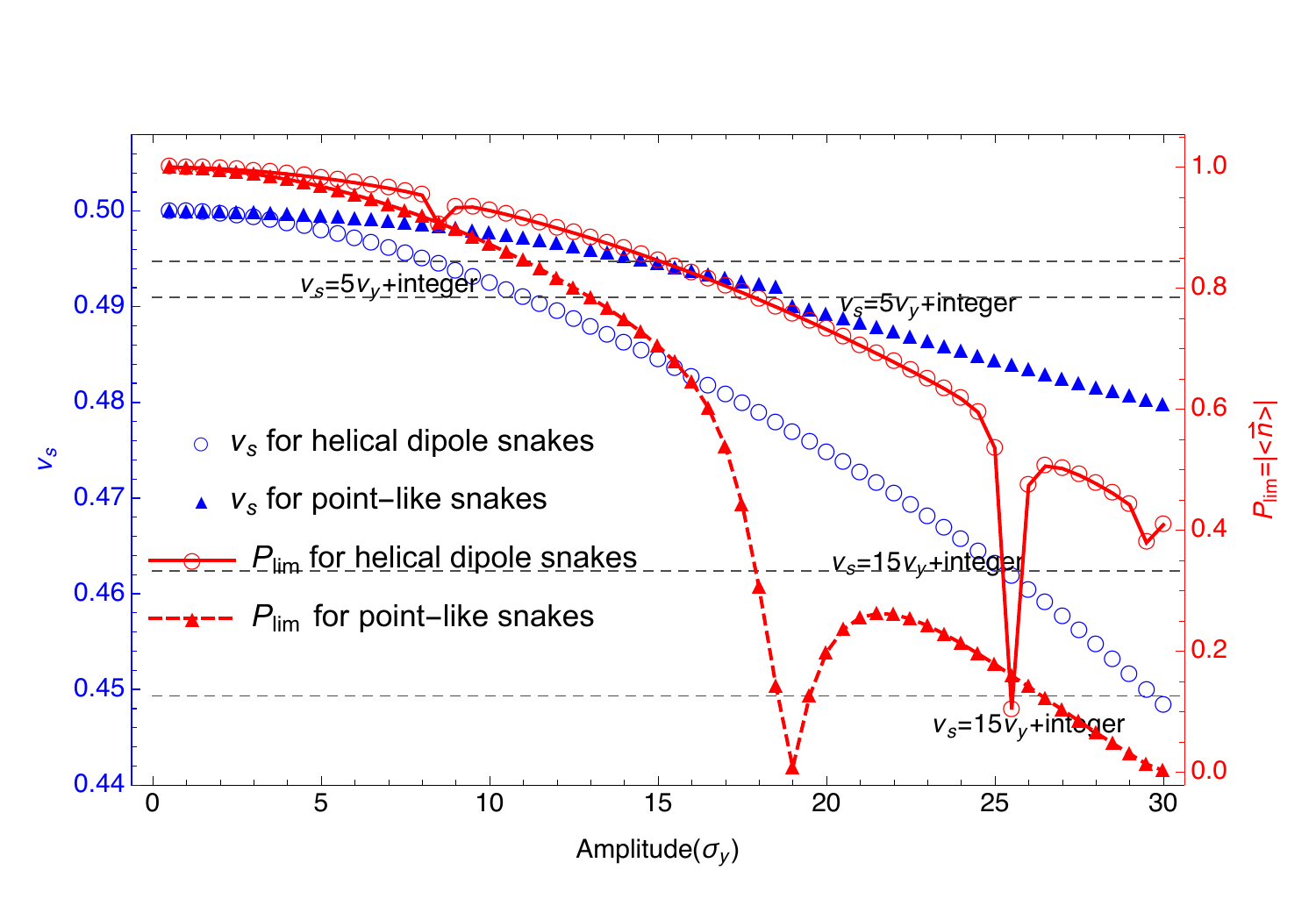}
        \caption{Case 1}
    \end{subfigure}%
    \begin{subfigure}{0.5\linewidth}
        \centering
        \includegraphics[width=75mm]{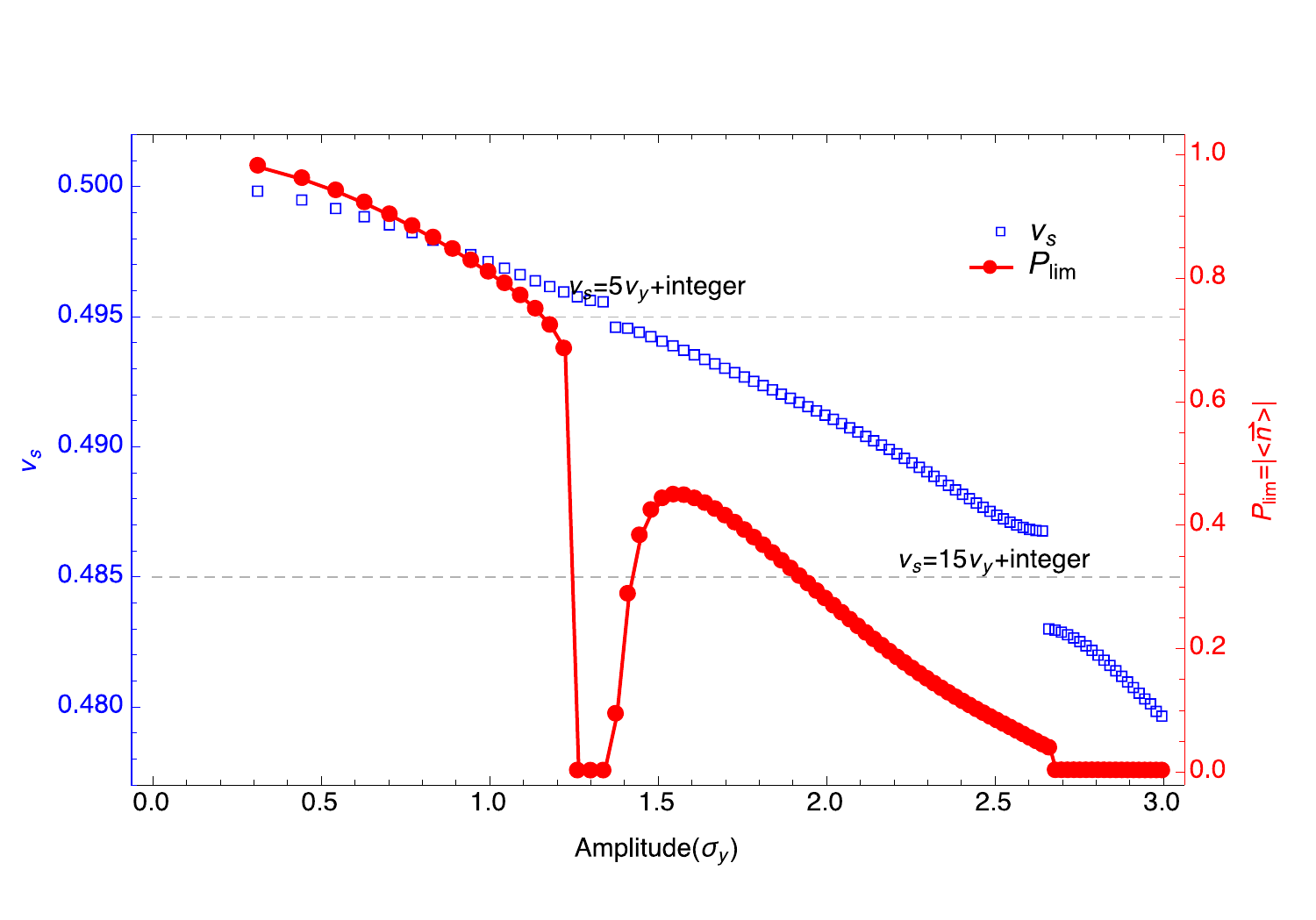}
        \caption{Case 2}
    \end{subfigure}
    \caption{(color online) $P_{\text{lim}}$ and $\nu_s$ versus
    vertical amplitudes for Cases 1 and 2. Here, the range of vertical betatron
    amplitudes is ten times larger in Case 1 than that in Case 2. 
    The betatron amplitudes are normalized by
    10$\pi~\text{mm}\cdot\text{mrad}$.
    The fractional vertical
    betatron tune near the closed orbit is set to 0.699.
    The helical dipole snakes are implemented for both cases, while the point-like snakes
    are also implemented for Case 1. In addition,
    the locations of visible higher order spin resonances are indicated as the intersections between
    the horizontal dashed lines and the $\nu_s$ curves in the plots. Note that there are two resonances of the form 
    $\nu_s=15\nu_y+\text{integer}$ in Case 1, due to the variation of $\nu_y$ with  such large
    vertical betatron amplitudes.}
    \label{fig:RHIC_ampscan}
\end{figure}

In Fig.~\ref{fig:RHIC_ampscan}, the behavior of $P_{\text{lim}}$ and $\nu_s$
is compared between Case 1 and Case 2,
and between different snake implementations.
$P_{\text{lim}}$ in general becomes smaller with increasing vertical betatron amplitude, and has 
dips near spin resonances.
Note that the range of vertical betatron amplitudes in Case 1 is 10 times larger than that in Case 2, 
and $P_{\text{lim}}$ decreases with amplitude much slower in Case 1 than Case 2,
with a much smaller underlying intrinsic resonance strength. 
In the lattice-independent model 
with a single vertical resonance driving term and two diametrically opposed orthogonal Siberian snakes, 
it was shown~\cite{Barber2003, S.R.Mane2008} that $\nu_s$ is 0.5 independently of the betatron amplitude,
if the fractional betatron tune is irrational. However, when the betatron amplitude becomes larger in the real lattice,
the nearby spin resonances are no longer isolated in both cases, 
and this analytical model is violated. 
Then we find that $\nu_s$ is shifted away from 0.5 with
amplitude, due to interference between nearby spin resonances.
Note that the shift of $\nu_s$ from 0.5 is in general much larger in Case 2 than Case 1 
for the same betatron amplitude. Moreover,
the locations of the spin resonances $5\nu_y-\nu_s=\text{integer}$ 
are indicated in the plots and match the sudden dips of $P_\text{lim}$ and 
the jump of $\nu_s$.
Several other higher order spin resonances are also visible and indicated in the plot,
and big jumps correlate with wide resonances.
In Case 1, the betatron amplitude becomes so large that the amplitude dependent orbital tune
shift is not negligible, and there are two locations corresponding to the same kind of spin resonance
$\nu_s=15\nu_y+\text{integer}$. In addition, in Case 1, except for the location of spin resonances,
$P_\text{lim}$ decreases faster with amplitude for the point-like snakes, while $\nu_s$ shifts faster away
from $0.5$ with amplitude for the helical dipole snakes, and the spin resonance $\nu_s=5\nu_y+\text{integer}$ is
wider for the point-like snakes. Because the helical dipoles also contribute
to the driving term of the spin resonances, for this case, it appears that the contribution from the helical dipoles 
cancels part of the total resonance strength mainly driven by the quadrupoles.

\subsection{Betatron tune scan \label{sec:tune}}

For a fixed vertical betatron amplitude, different vertical betatron tunes
correspond to different distances from major spin resonances. Therefore they
will lead to different values of $P_{\text{lim}}$ at the same vertical betatron
amplitude. In this section, the integer part of betatron tunes are kept constant so that
 ``vertical betatron tune'' refers to the fractional vertical betatron tune. 
A list of vertical betatron tunes in a selected range
is generated with a fixed step size, and the lattice is then fitted
accordingly for each case with a fixed fractional horizontal tune $\nu_x=0.69$. 
$P_{\text{lim}}$ is computed for these lattice
settings with the same vertical betatron amplitude 10$\pi~\text{mm}\cdot\text{mrad}$.
Helical dipole snakes are used in this simulation.

\begin{figure}[htb]
    \begin{subfigure}{0.5\linewidth}
        \centering
        \includegraphics[width=75mm]{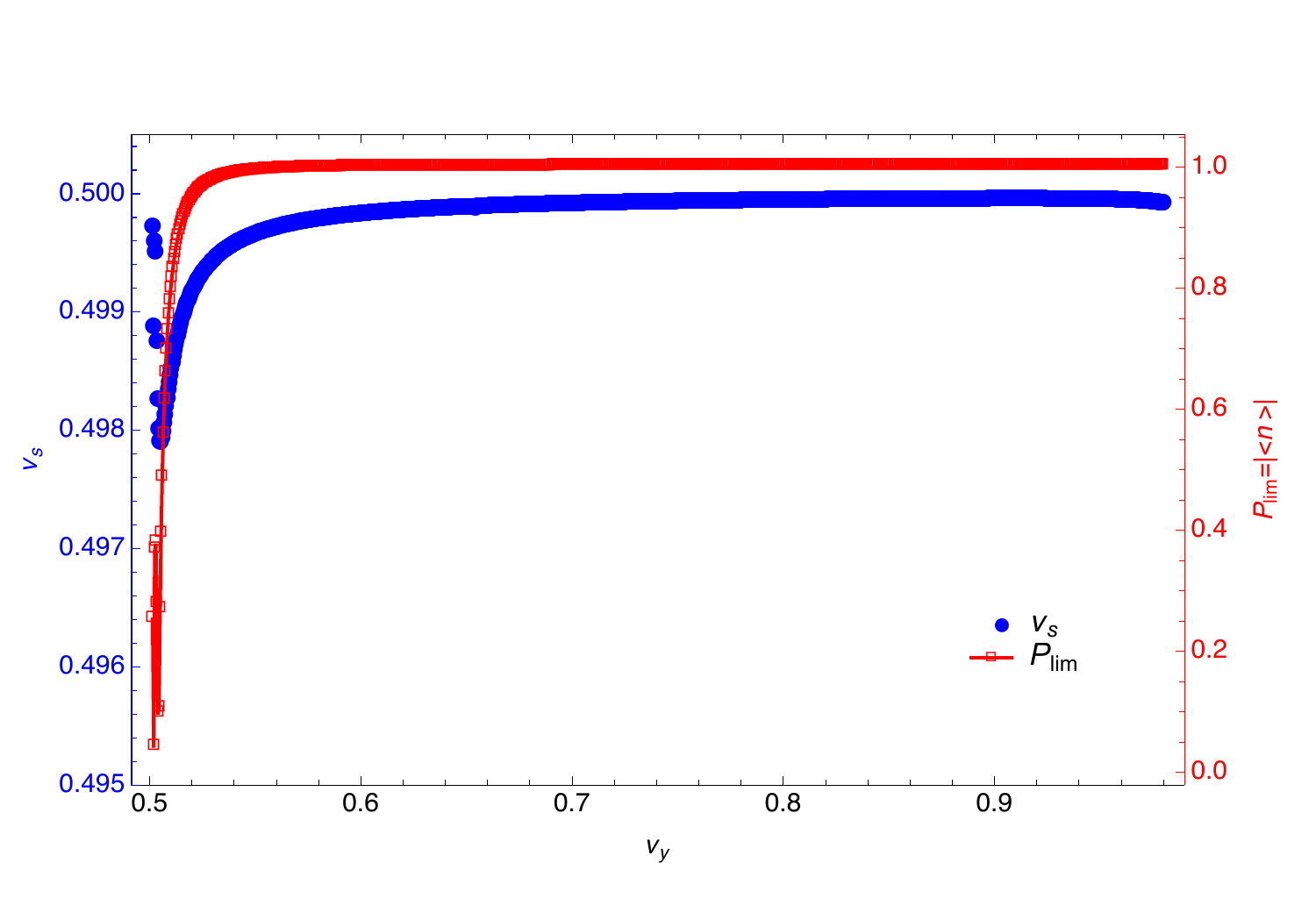}
        \caption{Case 1}
    \end{subfigure}%
    \begin{subfigure}{0.5\linewidth}
        \centering
        \includegraphics[width=75mm]{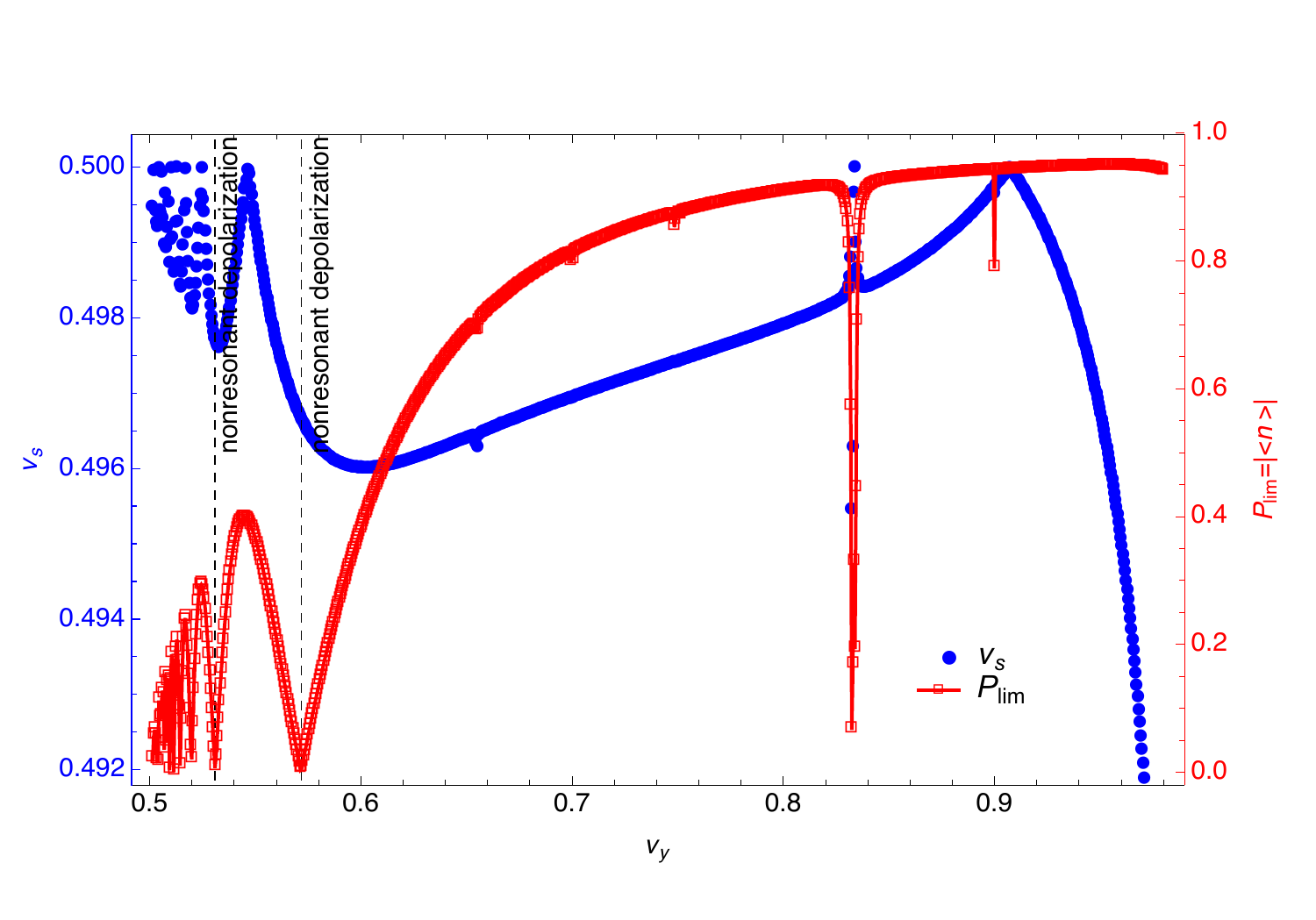}
        \caption{Case 2}
    \end{subfigure}
    \caption{(color online) $P_{\text{lim}}$ versus fractional vertical betatron tune in the range $[\nu_y]\in[0.501,0.980]$.  For both cases, the step size of tune scan is 0.0005. The vertical normalized emittance is 10$\pi~\text{mm}\cdot\text{mrad}$, and the horizontal tune is set to 0.69. Helical dipole snakes are implemented in both cases. Two locations of ``nonresonant beam polarization'' are indicated in Case 2.}
    \label{fig:tunescan1}
\end{figure}

Fig.~\ref{fig:tunescan1} is a scan in the range
$\nu_y\in[0.501,0.98]$ with a step size of 0.0005. 
It is clearly seen that in Case 2, $P_{\text{lim}}$ is
generally smaller when $\nu_y$ is closer to $0.5$, which 
indicates that the spin resonance $\nu_s=\nu_y+\text{integer}$ is so strong that it affects the behavior 
of $P_{\text{lim}}$ in the whole scan range.
In Case 1, however, the strength of the resonance $\nu_s=\nu_y+\text{integer}$ 
appears to be much smaller. 
Moreover, several other higher order spin resonances are also visible in Case 2, indicating that 
their widths are comparable or larger than the step size $0.0005$ in the vertical tune dimension.
Several locations of ``non-resonant beam depolarization'' are also observed in this plot, where 
the the dips in $P_{\text{lim}}$ do not correspond to a jump in $\nu_s$, i.e., the locations of spin resonances.
As shown in Ref.~\cite{S.R.Mane2004}, for the lattice-independent model with an isolated vertical 
resonance driving term and two diametrically opposed orthogonal snakes, $P_{\text{lim}}$ can be analytically 
expressed via a special function $a_0$
of the resonance strength, betatron tune and $G\gamma_0$, 
which goes to zero at the locations of
``non-resonant beam depolarization'' of that model.
This is an example where the study of these lattice-independent models
leads to physical insights that are nontrivial to obtain  otherwise.

\begin{figure}[htb]
    \begin{subfigure}{0.5\linewidth}
        \centering
        \includegraphics[width=75mm]{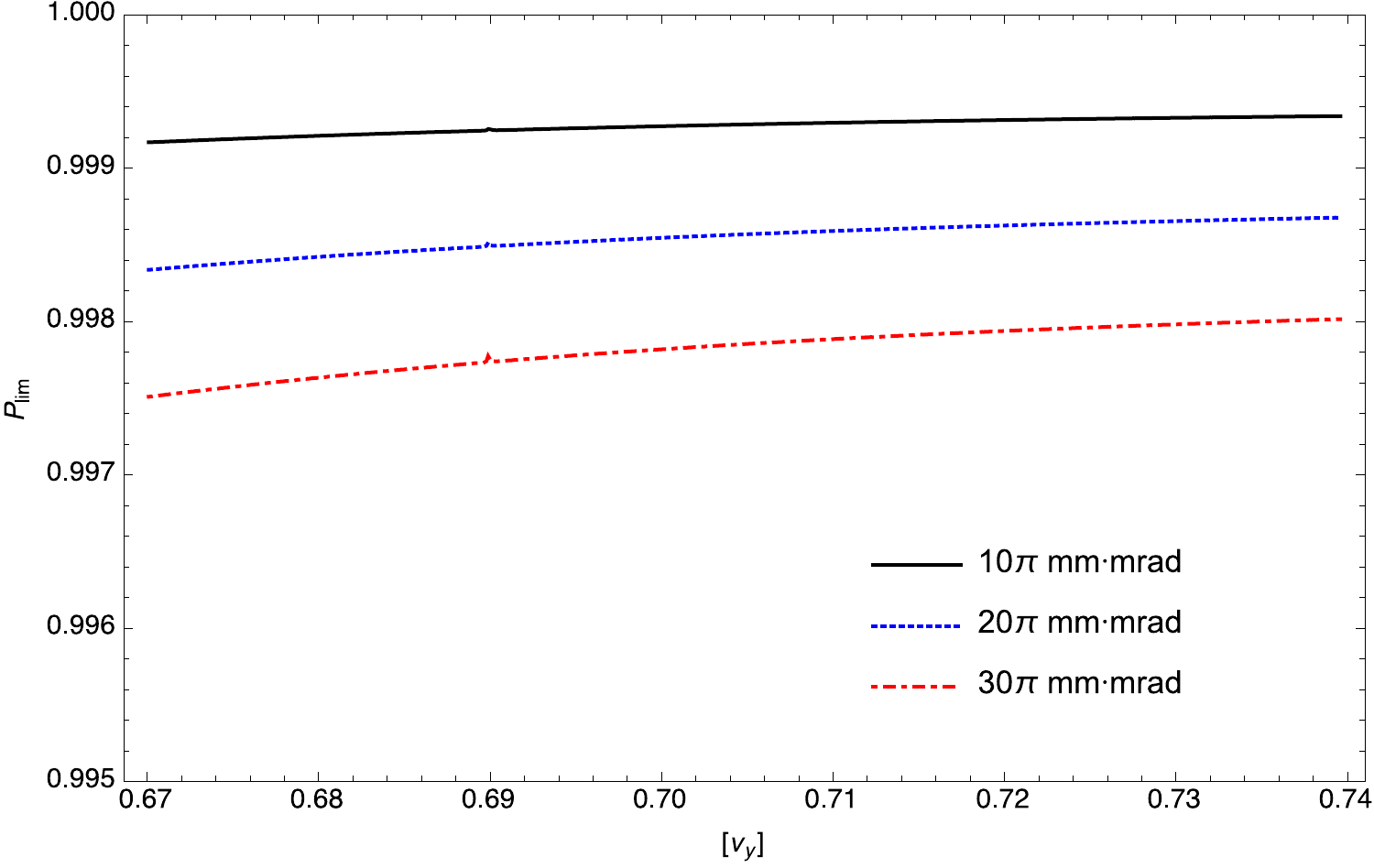}
        \caption{Case 1}
    \end{subfigure}%
    \begin{subfigure}{0.5\linewidth}
        \centering
        \includegraphics[width=75mm]{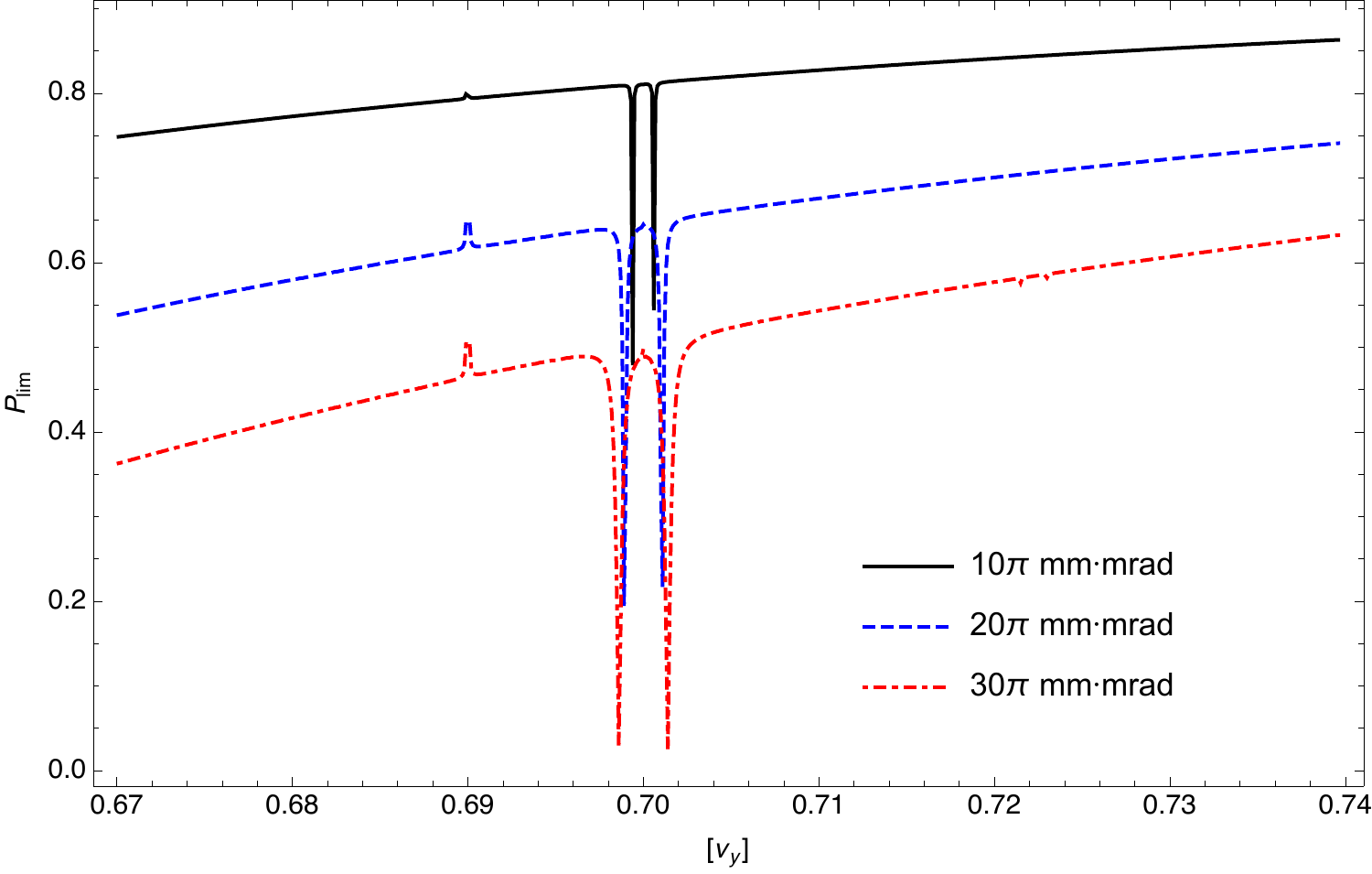}
        \caption{Case 2}
    \end{subfigure}
    \caption{(color online) $P_{\text{lim}}$ versus fractional vertical betatron tune in the range of $[\nu_y]\in[0.67,0.74]$ for both cases, the step size of the tune scan is 0.0001. Three different 
    vertical amplitudes are shown for Case 1 and Case 2. The horizontal tune is set to 0.69. Helical dipole snakes are implemented in both cases.}
    \label{fig:tunescan2}
\end{figure}

The vertical betatron tune range $[0.67,0.74]$ is of particular interests
because the vertical tunes of current RHIC operations are in this range. The
scan result with a step size $0.0001$ is shown in Fig.~\ref{fig:tunescan2},
where helical dipole snakes are applied. 
In Case 2, with this step size, 
it is clear that the
7/10 ``snake resonance'' is split into a doublet,
due to the fact that $\nu_s$ shifts with $\nu_y$. So the locations of the double resonances
shift with amplitude as well, while the widths of these spin resonances increase with amplitude.
Due to the transverse coupling introduced by the helical dipoles,
there is a small blip in $P_\text{lim}$ at $\nu_y=\nu_x=0.69$, which is
not visible if point-like snakes are used instead. However, in Case 1, the widths
of the spin resonances are very small and invisible in the plot.
Moreover, except for the resonance locations, 
it is shown that $P_{\text{lim}}$ increases with the vertical tune in this tune range. 

\subsection{Effect of the beam-beam interaction}
\begin{figure}[htb]
    \begin{subfigure}{0.5\linewidth}
        \centering
        \includegraphics[width=48mm]{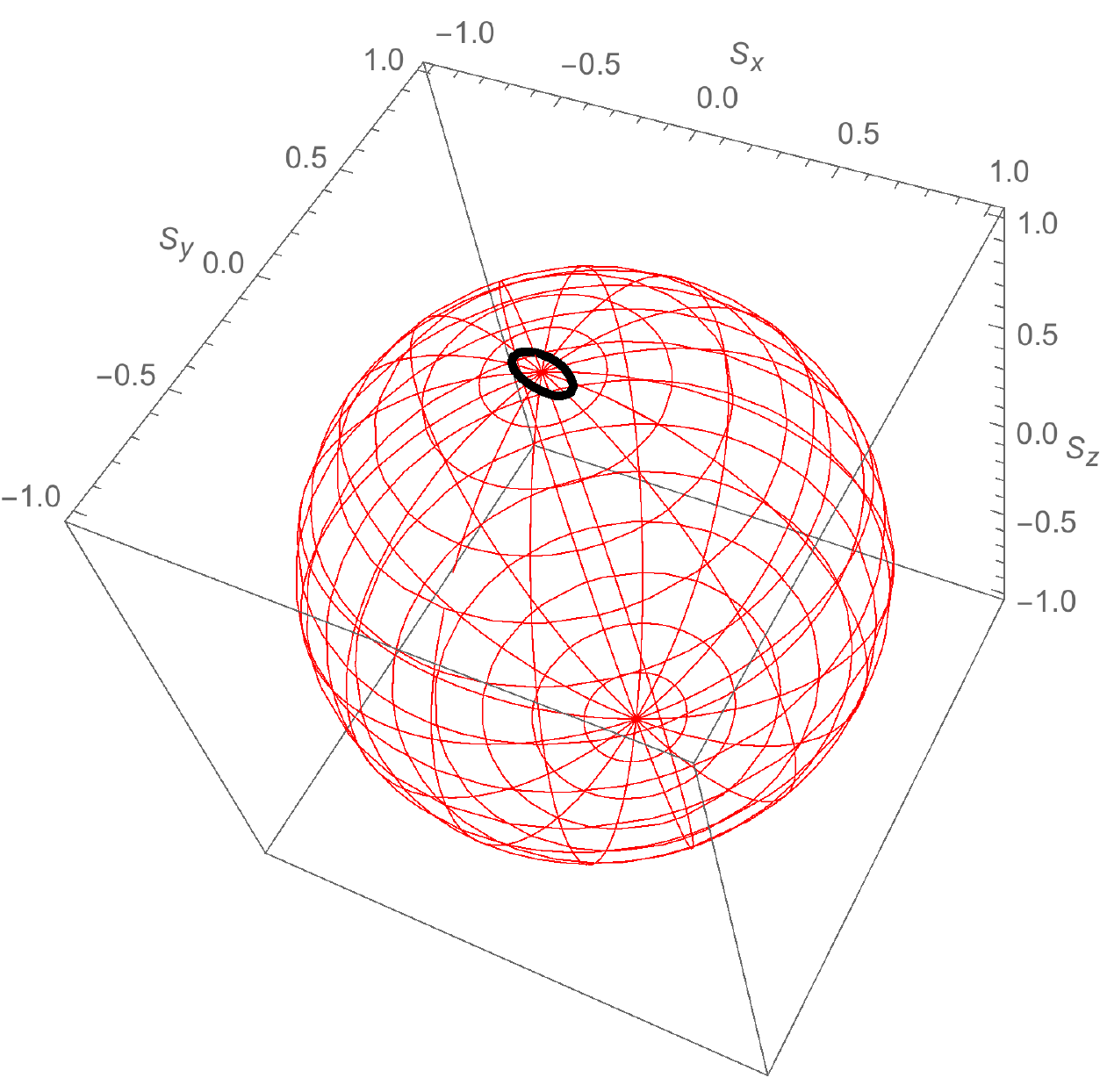}
        \caption{ The $\vec{n}$-axis on one phase space torus.}
    \end{subfigure}%
    \begin{subfigure}{0.5\linewidth}
        \centering
        \includegraphics[width=75mm]{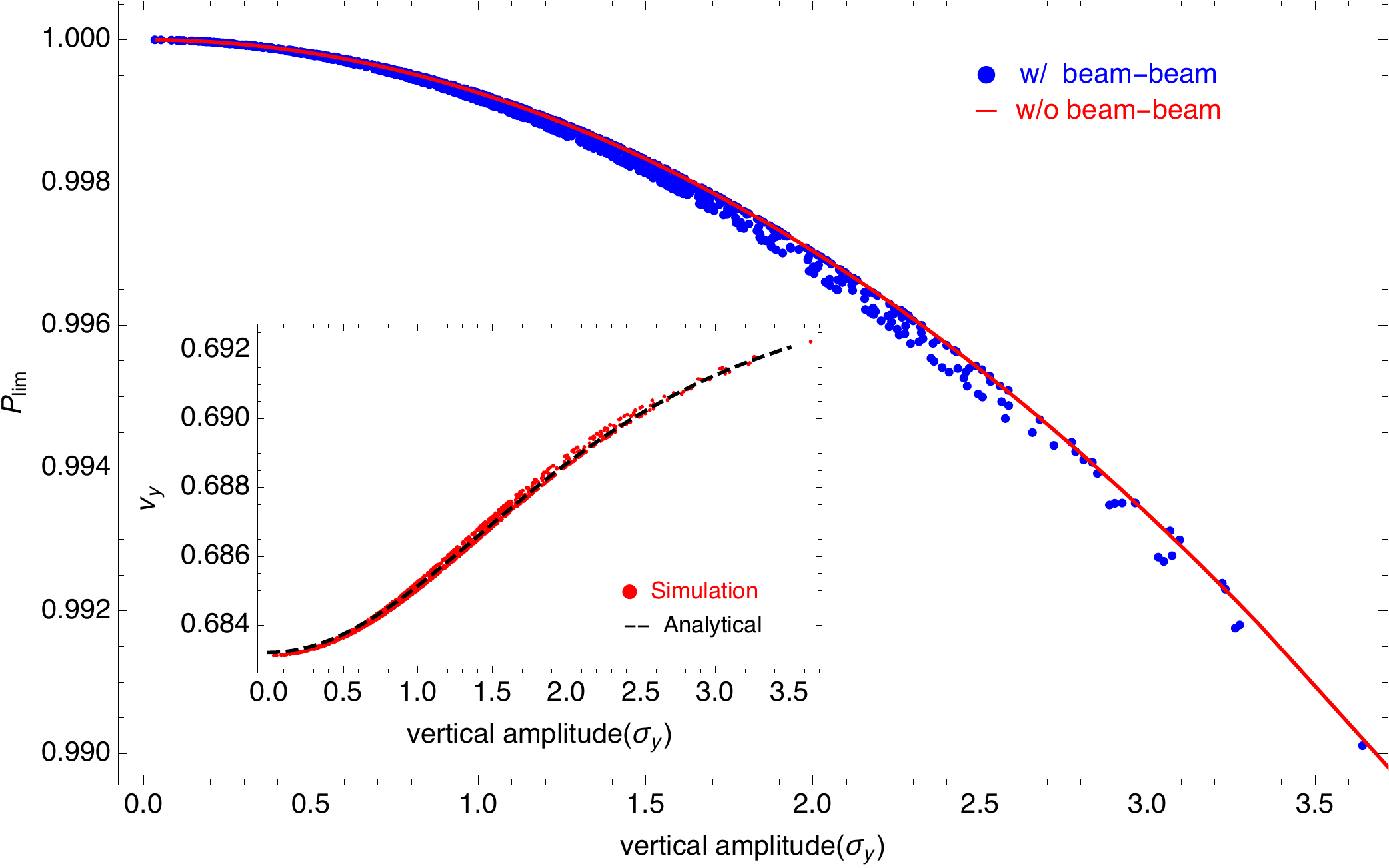}
        \caption{$P_{\text{lim}}$ and $\nu_y$ versus vertical amplitude}
    \end{subfigure}
    \caption{(color online) The effect of beam-beam interaction on $P_{\text{lim}}$.
    The left plot shows the turn-by-turn $\vec{n}$-axis of one tracked particle. The right
    plot shows $P_{\text{lim}}$ versus vertical amplitude for the cases with or without beam-beam
    interaction. In addition, the betatron tunes of tracked particles (``Simulation'') are compared with the analytical
    amplitude dependence of the tune shifts (``Analytical'')~\cite{Hirata}, in the presence of the beam-beam interaction.
    This is  shown in the small figure inside
    the right plot. 1008 particles with a Gaussian distribution were launched in the
    tracking. The beam-beam parameter is $-0.012$, and the fractional tunes are (0.690, 0.695).}
    \label{fig:BB}
\end{figure}

A thin-lens weak-strong beam-beam kick to the vertical betatron motion, is implemented at IP6,
with a vertical beam-beam parameter of $-0.012$. A beam of 1008 particles, 
with a Gaussian distribution for the  vertical coordinates
is launched and the $P_{\text{lim}}$ is computed for each particle's trajectory(torus). As shown
in Fig~\ref{fig:BB}, the calculated 
turn-by-turn $\vec{n}$-axis of one particle forms a closed curve on the surface of a unit sphere,
and this indicates the existence of an $\vec{n}$-axis on the particle's torus in the presence of
 nonlinear betatron motion.
Moreover, the effect of beam-beam interaction on $P_{\text{lim}}$ is insignificant. Note that the helical dipole 
snakes introduce a small transverse coupling in this example.

\subsection{Effect of horizontal motion}
\begin{figure}[htb]
    \begin{subfigure}{0.5\linewidth}
        \centering
        \includegraphics[width=75mm]{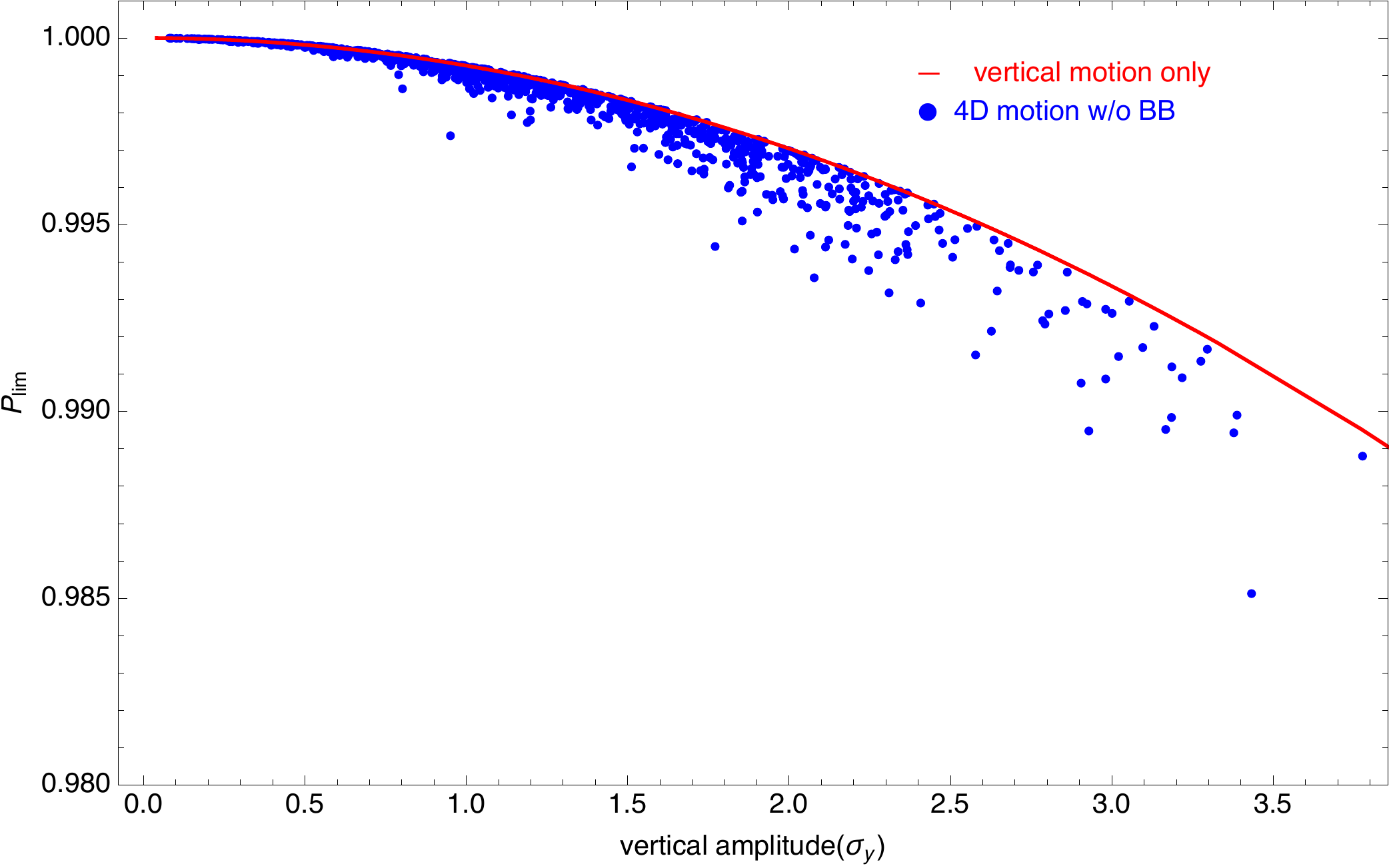}
        \caption{4D motion without beam-beam interaction}
    \end{subfigure}%
    \begin{subfigure}{0.5\linewidth}
        \centering
        \includegraphics[width=75mm]{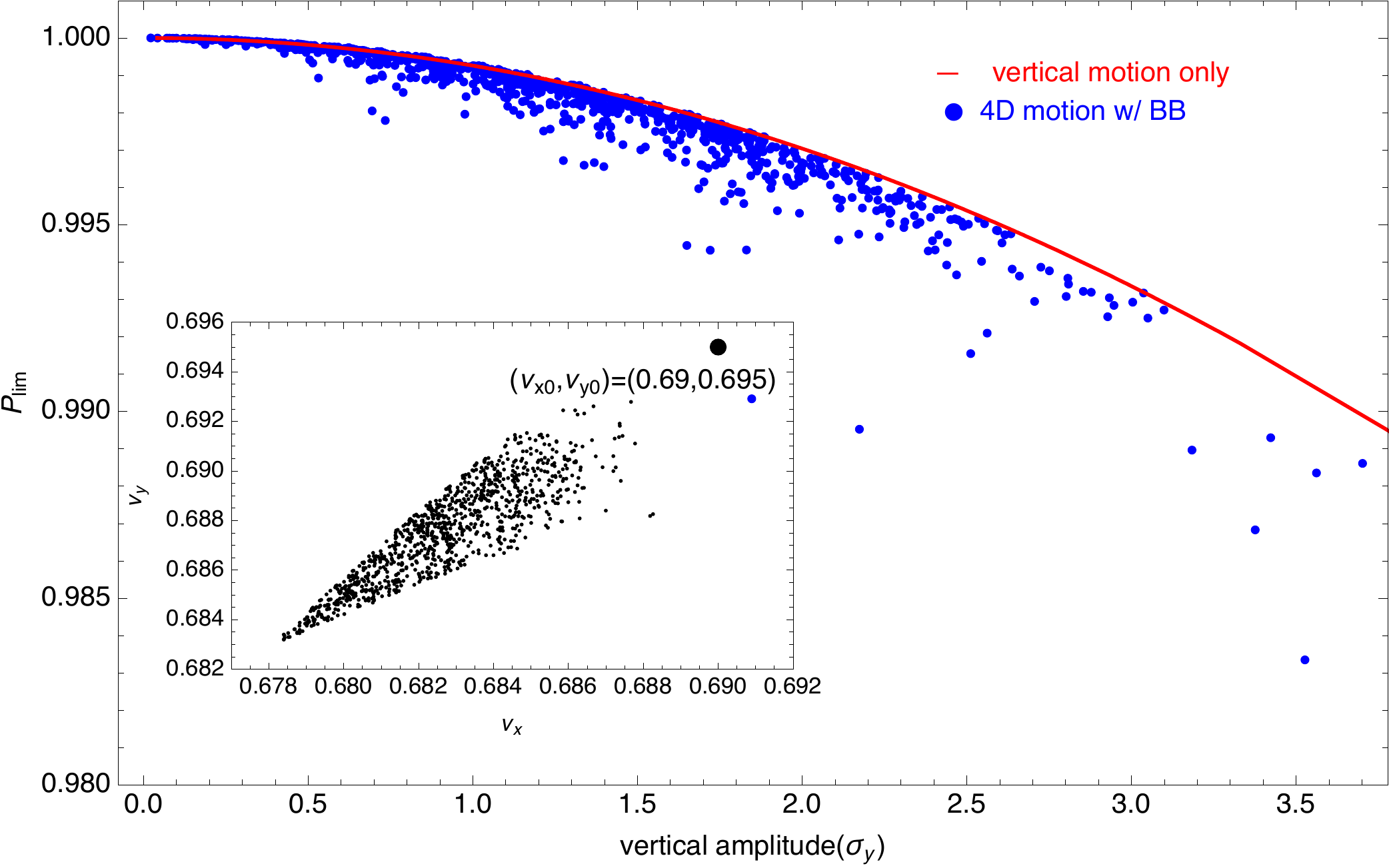}
        \caption{4D motion with beam-beam interaction}
    \end{subfigure}
    \caption{(color online) Comparison of $P_{\text{lim}}$ versus vertical betatron amplitude, as normalized by $10\pi\text{mm}\cdot\text{mrad}$ for 1008 trajectories, between the case with 4D motion, and the case with only vertical betatron motion. The left plot shows the case without beam-beam interaction, and the right plot shows the case with beam-beam interaction. The 95\% normalized emittances are 10$\pi~\text{mm}\cdot\text{mrad}$ for both horizontal and vertical planes. A beam-beam kick with a beam-beam parameter -0.012 is implemented at IP6 of the RHIC lattice, for the right plot, and the tune footprint of the beam particles is shown in a small figure inside
    the right plot as well. The fractional tunes are set to $(0.69, 0.695)$.}
    \label{Fig:4DPLIM}
\end{figure}

The simulations for the cases shown above deal primarily with the vertical betatron motion, since the contribution of
the small horizontal amplitude introduced by the helical dipoles is small for these cases.
We now include horizontal motion 
by initializing a beam with a 4D Gaussian distribution.
The normalized 95\% emittances are 10$\pi~\text{mm}\cdot\text{mrad}$ for both
horizontal and vertical planes.
 Two cases are simulated as shown in Fig.~\ref{Fig:4DPLIM}, one without
 beam-beam interaction, and the other with a beam-beam kick at IP6, whose
 beam-beam parameter is -0.012.
Compared to the case with
only vertical betatron motion, when 4D motion is included, $P_{\text{lim}}$
spreads out with vertical amplitude to some extent, for different trajectories. However, the two
cases with or without beam-beam effect do not show much difference. This is 
because the variation of $P_{\text{lim}}$ with vertical betatron tune is very small for Case
1 in the tune range between 0.67 and 0.74, as shown in
Fig.~\ref{fig:tunescan2}.

\section{Conclusion}
In this paper we compute the static polarization limit for a practical RHIC
lattice with the physics store conditions for various beam parameters, as a
step towards understanding the polarization evolution at store. All
calculations are done on the basis of the Polymorphic Tracking Code. 
It is shown
that the vertical betatron oscillation has the dominant effect on the behavior of the $P_{\text{lim}}$
 in contrast with the horizontal oscillations. Note that synchrotron 
motion is not included in this simulation because the synchrotron tune of RHIC is very small, namely
around $3\times10^{-4}$ at store.
In this case the use of stroboscopic averaging to find the $\vec{n}$-axis when synchrotron motion is included requires special studies.
Moreover,
the practical
modeling of Siberian snakes with helical dipoles leads to different behavior of $P_{\text{lim}}$
and $\nu_s$, in contrast to the implementation of the point-like Siberian snakes. So it is advisable to model the snakes carefully.
 The ``nonresonant beam polarization'' observed
and studied in the lattice-independent model is also observed in this lattice-dependent model.
Moreover, the beam-beam interaction doesn't have much effect on $P_{\text{lim}}$ for the 
parameters under study. In addition, machine imperfections can possibly 
tilt $\vec{n}_0$ from the vertical and thereby lead to spin resonances due to horizontal motion.
Imperfections can also 
shift $\nu_0$ away from 0.5~\cite{Ptitsyn:2010zz}.
A realistic treatment of various sources of machine imperfections requires a very careful
lattice modeling, and is beyond the scope of this paper.
Nevertheless, this work shows how a study of $P_{\text{lim}}$
and $\nu_s$ can give insights at fixed energies that are not
available by executing simple tracking. 

This work is supported under Contract
No. DE-AC02-98CH10886 with the U.S. Department of
Energy, the Hundred-Talent Program (Chinese Academy
of Sciences), and National Natural Science Foundation of China
(11105164). 
We would like to thank Drs. D. Abell and E. Forest on
their help with the simulation code PTC,   
Drs. M. Bai, D. P. Barber, F. Meot, Y. Luo,
V. Ptitsyn, V. Ranjbar and T. Roser for helpful discussions.   
One of us, Z. Duan, would like to thank Prof. M. Bai for being his host during his stay in BNL,
guiding him into this field and the many instructive discussions over the years. He also would like
to thank Dr D. P. Barber for helpful suggestions and discussions and careful reading of the manuscript.
The simulation work used the
resources of the National Energy Research Scientific Computing
Center, a DOE Office of Science User Facility supported by the Office
of Science of the U.S. Department of Energy under Contract No.
DE-AC02-05CH11231.

\end{document}